\documentclass[reprint,superscriptaddress,aps,prl]{revtex4-1}
\usepackage{graphicx}
\usepackage{xcolor,color}
\usepackage{amsmath} 
\usepackage{amsthm} 
\usepackage{amssymb}	
\usepackage{hyperref}
 
\newcommand{\ang}{\mathrm{\AA}}
\newcommand{\ket}[1]{\left| #1 \right>} 
\newcommand{\bra}[1]{\left< #1 \right|} 
\let\baraccent=\= 

\begin{document}

\title{Looking Beyond the Surface with Angle-Resolved Photoemission Spectroscopy}

\author{R. P.\,Day}
\email[]{rpday7@gmail.com}
\author{I. S.\, Elfimov}
\author{A.\,Damascelli}
\email[]{damascelli@physics.ubc.ca}
\affiliation{Department of Physics $\&$ Astronomy, University of British Columbia, Vancouver, BC V6T 1Z1, Canada}
\affiliation{Quantum Matter Institute, University of British Columbia, Vancouver, BC V6T 1Z4, Canada}

\date{\today}

\begin{abstract}

The issue of surface sensitivity, and its relationship with interpretation of spectral features observed in angle-resolved photoemission spectroscopy experiments is investigated. Rather than attempt to make an explicit connection to bulk electronic structure calculations, we take the new approach of exploring this issue within the natural context of a vacuum-terminated crystalline slab. Doing so, we reconcile the empirical reality of reliable $k_z$ fidelity with acute surface sensitivity of this technique. In addition, we identify several critical issues which impact the estimation of $k_F$, band velocity, and  self-energy from photoemission experiments.
\end{abstract}
\maketitle

\section{Introduction}

Angle-resolved photoemission spectroscopy (ARPES) has played an increasingly important role in the study of quantum materials over the past two decades \cite{Hufner1995,Dama2004,Sobota2020}. In particular, ARPES offers a direct measure of the Fermi surface and valence band electronic structure, without reliance on approximate or phenomenological models. Such an experimental description of the electronic structure provides explicit insight on transport properties and various thermodynamic probes, and informs the construction of realistic theoretical models of material phenomenology. 
Although the surface sensitivity of photoemission has its merits  \cite{Stampfl_1995, Hossain_2008, Hsieh_2009,Tan2013}, it is also at the root of a recurring criticism of the technique's reliability; while alternatives such as quantum oscillations probe the bulk electronic structure, the mean free path of photoelectrons excited with ultraviolet radiation renders photoemission sensitive to only a few unit cells nearest the vacuum interface \cite{Seah1979}. The questions of surface quality, termination, polarity, and reconstruction make the effort to study bulk physics via ARPES a formidable challenge in many materials of interest \cite{Dama2000,Yang2010,Mazzola2018,Cucchi2019,Hideaki2019, Sato2020}. For these reasons, only in those quasi-two dimensional materials with very weak interlayer coupling is the surface to be expected to be representative of bulk physics. \textit{To what extent, however, are these concerns warranted, and when can one rely on photoemission as a suitable probe of bulk properties?} These are the questions we seek to answer in this work.

\begin{figure}
	\includegraphics[width=\columnwidth]{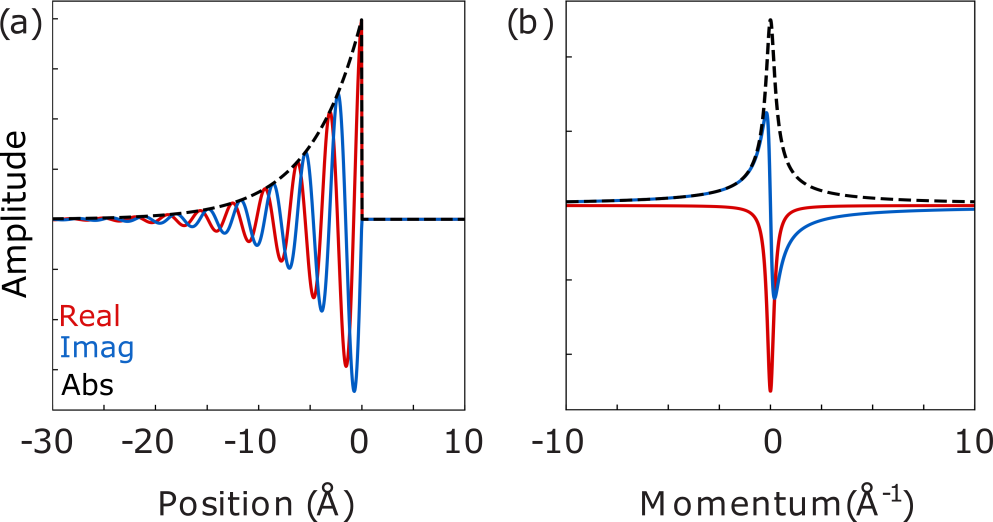}
	\caption{Surface sensitivity as the source of momentum uncertainty. The Lambert relation $f(z) = f_0e^{(ik_z^0 - 1/\lambda)z} $ plot in (a) approximates the photoelectron's depth-projected wavefunction envelope, which under Fourier transformation translates to a Lorentzian $k_z$ distribution with width $1/\lambda$, as in panel (b).} \label{fig:fourier}
\end{figure}

The surface sensitivity of ARPES is commonly introduced as a consequence of the finite escape depth, or mean-free-path ($\lambda$), of the high kinetic-energy photoelectrons; those electrons which scatter between photoemission and detection cannot readily be mapped onto the electron-removal spectral function \cite{Pendry1976,Lee1999}. This is alternatively described as relating $\lambda$ to the imaginary part of the crystal's inner potential, the electronic analogue to the complex index of refraction \cite{Hedin2001,Minar2014}. A suitable description of the photoemission intensity which accounts for a finite $\lambda$ can be written as:
\begin{equation}\label{PES}
\begin{split}
I(k,\omega) =& \sum_{i,f}|\langle N-1,f; \tilde{k}|\Delta c_k | N,i\rangle|^2\times\\
&\delta(h\nu - [E_{N-1}^f + \omega - E_{N}^i]),
\end{split}
\end{equation}
where $\tilde{k} = k + i/2\lambda$. The initial state $|N,i\rangle$ refers to the initial $N$-body wavefunction, and $|N-1,f;\tilde{k}\rangle$ to all possible $N-1$ particle final state accompanied by a free-particle state with wavenumber $\tilde{k}$ excited via photoemission. The delta-function imposes energy conservation. In what follows here, we make the specific choice to describe the photoelectron final state as a damped plane-wave. While a more sophisticated approach may utilize KKR-scattering final states  \cite{Minar2014}, this approximation simplifies the discussion without loss of generality for our conclusions. Doing so, we rewrite Eq. \ref{PES} as:
\begin{equation}\label{PES_spectral}
I(k,\omega) = \sum_{i,f} |\left < e^{ik\cdot r}e^{-z/2\lambda} | \Delta |\phi_i\right>|^2 A_{i,f}(k,\omega).
\end{equation}
The electron-removal spectral function is encapsulated within $A_{i,f}(k,\omega)$, and the dipole interaction matrix element has been expressed in a basis which includes a damped outgoing plane wave and the single-particle states $\{\phi_i\}$ appropriate to a description of the material under study. The exponential attenuation factor in the photoelectron final state captures the essence of the surface sensitivity. Ordinarily, a connection is made between this exponential factor, and the ability to measure distinct $k_z$-specific electronic structure via ARPES. In fact, one can interpret this exponential localization as a position-measurement, from which the experimentalist gains detailed information regarding the origin of those photoelectrons which are ultimately collected in the spectrometer. As such, one can make an uncertainty-principle argument, from which we arrive at the conclusion that $\Delta k_z\sim 1/\lambda$. This can be made more rigourous, applying a Fourier transform to the Lambert relation $f(z)=f_0e^{-z/{\lambda}}$ to obtain:
\begin{equation}\label{eq:FT}
F(k_z) = f_0 \int_{-\infty}^{0} dz\ e^{(i(k_z-k_z^0)  - 1/\lambda)z} = \frac{1}{i(k_z-k_z^0) - 1/\lambda}.
\end{equation}
Treating the $|F(k_z)|^2$ as a power spectrum over $k_z$, we recover a distribution with spectral width of $1/\lambda$, as visualized in Fig. \ref{fig:fourier}. This uncertainty-argument then arrives at the same $k_z$ distribution which has been argued on a different basis previously  \cite{Chiang1998,Hedin2001,Strocov2003}. 
From the values of $\lambda$ predicted for photon energies in the range of 20-100 eV relevant to most ARPES experiments \cite{Seah1979,Hofmann_2002}, one would infer that $k_z$-specific ARPES is fundamentally untenable. Over this domain, $\lambda$ is of the same order as the $c$-axis parameter for most materials. Consequently, $\Delta k_z$ would subsume the entire out-of-plane axis of the Brillouin zone, integrating over any three-dimensional dispersion indiscriminately. However, it is a matter of experimental record that high-fidelity $k_z$-dependent studies of the three-dimensional dispersion of many materials have been achieved, often in exceptional agreement with conventional bulk-probes like quantum oscillations \cite{Dama2000}. We conclude then that something is lacking in this simple description of surface sensitivity in photoemission. 

In the following, we introduce the role of the surface explicitly, modelling both the electronic structure and the photoemission process within the same surface-specific context. Doing so, we identify a simple explanation for the unexpectedly high $k_z$-fidelity achieved via UV-ARPES experiments, while also revealing several important caveats which warrant serious consideration in regards to how we interpret the spectroscopic features observed in ARPES. Specifically, we will focus on the reliability with which we can extract band positions ($\epsilon_k$), and the electron self-energy ($\Sigma$) or quasiparticle scattering rate from the intensity lineshapes. 

\section{Photoemission at the Interface}
Recently, we have applied the methodology described here in a multi-orbital iron-based superconductor \cite{Day_2021}. Presently however, in pursuit of ultimate transparency, we will use a simple tight-binding model, defined with a basis of $s$-orbitals on a cubic lattice of dimension $a=5$\AA. We restrict kinetic terms to nearest-neighbour hopping only, with amplitude $t$. This results in a single energy scale $t$, as $t=t_{||}=t_{\perp}$. In connection to the conventional ARPES geometry, we use $\perp$ and $z$ or $c$ axis interchangeably, reserving the parallel directions to be those lying within the surface plane of the crystal. We note that as the reference energy, or Fermi level is chosen arbitrarily in this model, where we have set it sufficiently high that the Fermi function does not directly interfere with our analysis of the spectral function.  In contrast to this simple model Hamiltonian, it is important to emphasize that we often contend with more complicated geometries in real materials, with multiple hopping energy scales $\{t_i\}$  depending on orbital character and inter-atomic vectors \cite{Slater_1954}. The consequences for surface sensitivity become even more significant in those scenarios. For example, in the last section of this paper, we will explore the implications for $t_{\perp}\neq t_{||}$. 

\begin{figure}
	\includegraphics[width=\columnwidth]{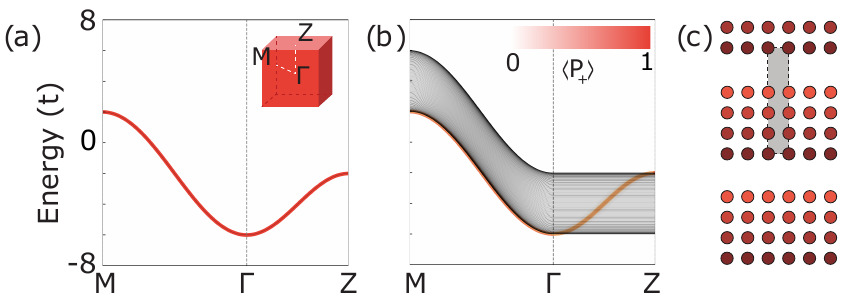}
	\caption{Cubic lattice model. The bandstructure is plot along high symmetry directions of the bulk Brillouin zone for bulk (a) and 100-layer slab (b) models. The expectation value of the projector $P_+$ is plot over each band, varying from transparent (0) to red (1).  In (c), a schematic of the slab unit cell for 4-layers is indicated, with the unit cell identified by the grey volume.} \label{fig:model}
\end{figure}

The model Hamiltonians for this work have been constructed using the $chinook$ software \cite{Day2019}. The bandstructure of the bulk model is plot in Fig. \ref{fig:model}(a). To introduce the crystalline surface, we utilize a vacuum-terminated slab geometry: after extending the unit cell along the (001) direction, several layers are deleted, introducing a large vacuum layer within the unit cell \cite{Sun_2013}. Doing so, we construct a finite crystal with a vacuum interface, while formally preserving periodic-boundary conditions (PBC) along all three principal axes of the unit cell. The vacuum layer is chosen sufficiently thick to suppress any inter-unit cell hopping along the $k_z$ direction, similar to imposing open boundary conditions (OBC) along this direction. In this work, we use a 200-layer slab with a 10\AA vacuum buffer. Although this heterostructure resembles the case of a free-standing finite slab of material, our unit cell is a somewhat artificial construction. In particular, it preserves the concept of a $k_z$ quantum-number, inherited from the periodic nature of the model. In the Appendix, we make a direct connection between PBC and OBC: ultimately, the two are compatible. The $k_z$ information encoded in the electronic initial state wavefunctions for PBC carries over to the photoelectron final state for OBC. By virtue of the computational facility with which we can solve the problem, we proceed here with the slab geometry. 

For a unit cell with $m$ basis orbitals, the corresponding slab with a thickness of $N$ unit cells is characterized by an eigenspectrum with $mN$ levels. These energy bands are non-dispersive along the $k_z$ direction. In such a slab geometry, one may observe a combination of slab-delocalized quantum well states, and surface-localized states such as Tamm, Shockley, and topological surface states \cite{Harrison2003,Fu2007,Hsieh_2009,Tamai2013}. In the model we consider here, the symmetry breaking potential of the vacuum is insufficient to elicit surface-localized states. The bandstructure of the slab unit cell is plot in Fig. \ref{fig:model}(b) along the same path as the bulk model in Fig. \ref{fig:model}(a). In the limit $N\rightarrow \infty$, the bandstructure forms a continuum spanning the bulk-model's $k_z$-bandwidth, with an indistinguishable density of states \cite{Starfelt_2018}. 

\begin{figure*}
	\includegraphics[width=\textwidth]{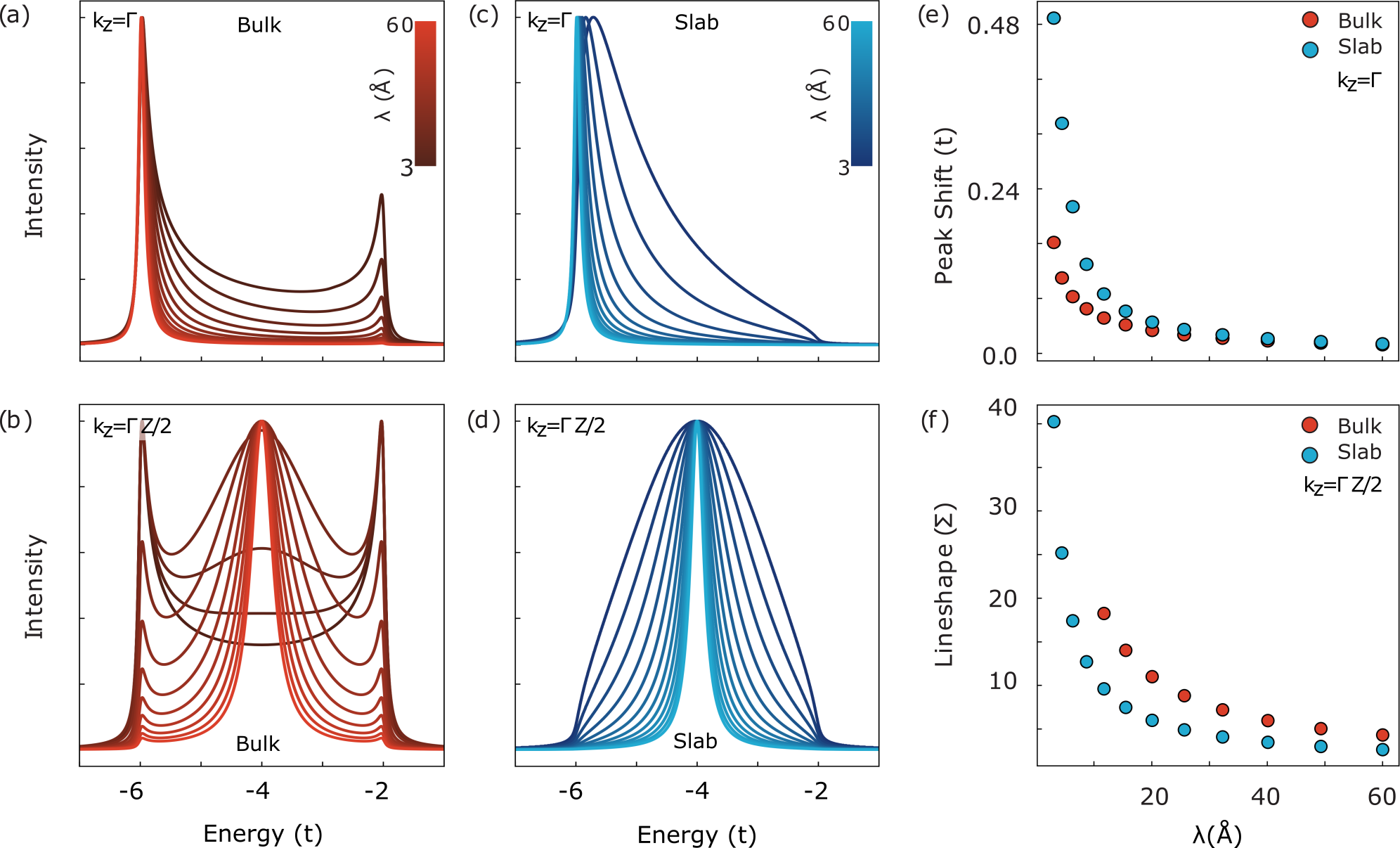}
	\caption{Slab vs. bulk model comparison for spectral analysis. In (a-d), we plot simulated ARPES EDCs at the $k_z=\Gamma$ (a,c), and $k_z=\Gamma Z/2$ (b,d) points, as computed using the bulk (a,b) and slab (c,d) methods described in the text. Each spectrum is computed for a range of photoelectron escape depths from $\lambda=3$ to $60\ang$, indicated by the red and blue colourscales in (a,c) for bulk and slab, respectively. At the band edge ($k_z=\Gamma$), a distinct shift in the peak position is observed, specifically for the slab calculation. This is made explicit with the peak shifts plot in (e), as measured relative to the bulk band energy. Similarly, near the band centre ($k_z=\Gamma Z/2$) the spectra are amenable to Lorentzian fitting in the vicinity of the peak, enabling estimation of the apparent peak lineshape (HWHM), as plot in (f) in units of the intrinsic imaginary electron self energy ($\Sigma$). Alll spectra in (a-d) are renormalized by their peak intensities to enable direct comparison.} \label{fig:spectra}
\end{figure*}

While the level structure lacks any $k_z$-dependence, the use of PBC imbues the eigenvectors with $k_z$ information. This can be illustrated by plotting the symmetric projection operator. We define:
\begin{equation}\label{eq:proj}
P_+ = \frac{1}{N}\sum_{\alpha}\ket{\phi_{\alpha}} \bra{\phi_{\alpha}},
\end{equation}
where $\ket{\phi_{\alpha}}$ is the basis state located in the $\alpha$-th layer of the slab. In the infinite lattice, the eigenstates are Bloch waves, which amount to a symmetric linear combination over Wannier functions centred at the atomic-site in each unit cell. This projector $P_+$ then serves to illuminate vestigial signatures of the infinite crystal's Bloch states within the finite slab's electronic structure. Perhaps unsurprisingly, $\left< P_+ \right>$ is finite only close to the dispersion of the bulk Hamiltonian. Away from the points where $\left< P_+ \right>\neq 0$, the eigenstates are various orthogonal linear combinations over the slab, all of which are functions of $k_z$. Although the energy spectrum is $k_z$ independent, we see that the dispersive Bloch wave of the bulk model is largely preserved, woven into the continuum of slab-eigenstates. 

Equipped with this slab-Hamiltonian, we are prepared to discuss the $k_z$-sensitivity and fidelity in ARPES measurements in the small $\lambda$ limit. One can take two different approaches to emulating the role of surface-sensitivity in the photoemission experiment. In the first, we apply the Lorentzian distribution from Eq. \ref{eq:FT} to reflect the bulk model:
\begin{equation}\label{eq:I_FT}
I_{FT}(k_{||},\omega) = \sum_{k_z} |F(k_z)|^2 I(k_{||},k_z,\omega).
\end{equation}
Throughout, we will make the simplifying assumption that the electron-removal spectral function $A_{i,f}(k,\omega)$ is diagonal in the band basis, and takes the functional form of a Lorentzian with half width half max (HWHM) $\Sigma$, independent of energy, momentum, and orbital or band index. Combining this with Eq. \ref{eq:FT}, we can rewrite Eq. \ref{eq:I_FT} as:
\begin{equation}
\begin{split}
I_{FT}(k_{||},\omega) =\sum_{i=1}^{m} \sum_{k_z}& \frac{\frac{1}{\lambda}}{(k_z-k_z^{0})^2 + (\frac{1}{\lambda})^2}\\
&\frac{\Sigma}{(\omega-\epsilon_{bulk}^i)^2 + (\Sigma)^2}|\left <e^{ik\cdot r}|\Delta|\psi_i\right>|^2.
\end{split}
\end{equation}
Here $k_z^{0}$ is the $k_z$ value at which the distribution is centred, and $\epsilon_{bulk}^i$ the bulk Hamiltonian eigenenergy for eigenstate $|\psi_i\rangle$. Alternatively, we can simulate ARPES intensity directly from the slab-model. To do so, we express the intensity as:
\begin{equation}
I_{S}(k_{||},\omega) =\sum_{i=1}^{mN}  
\frac{\Sigma}{(\omega-\epsilon_{slab}^i)^2 + (\Sigma)^2}|\langle e^{ik\cdot r-z/2\lambda}|\Delta|\psi_i\rangle|^2.
\end{equation}\label{eq:I_slab}
The eigenstates $|\psi_i\rangle$ and eigenenergies $\epsilon_{slab}^{i}$ refer to the slab Hamiltonian explicitly. In this latter case, we introduce surface-sensitivity by damping the final-state plane waves, applying an exponential extinction factor to each basis-state's contribution to the photoemission cross-section. By comparing and contrasting the two approaches, we aim to understand why $k_z$-dependent ARPES is possible, even in the small $\lambda$ limit relevant to mid-ultraviolet excitation energies. 

 In Fig. \ref{fig:spectra}, we plot results under both schemes, for various values of $\lambda$ at $k_z = 0$ and $k_z = \pi/2$. While each converges to the intrinsic lineshape (HWHM) and binding energy of the bulk Hamiltonian in the large $\lambda$ limit, there are several important differences in the range relevant to UV-ARPES experiments. In regards to both lineshape and peak position, there is a failure to reproduce the intrinsic spectral function lineshape and binding energies associated with the bulk Hamiltonian as $\lambda$ approaches zero.

Perhaps surprisingly, although the eigenspectrum of the slab-model is a non-dispersive continuum, the photoemission spectra are peaked near the associated bulk band locations for each $k_z$. The peak position, plot for example for $k_z=0$ in Fig. \ref{fig:spectra}(e), converges rapidly to the bulk value with increasing $\lambda$. By contrast with this, the $k_z$-integrated bulk simulation integrates indiscriminately over the $k_z$-projected density of states as $\lambda \rightarrow 0$, resulting in peaks at the band edges. 

In addition to the peak position, the peak lineshape also evolves with $\lambda$. The characteristic lengthscale over which the lineshape of $I_{FT}$ decays is approximately twice that of the vacuum-terminated slab calculations, as illustrated in Fig. \ref{fig:spectra}(f). Nonetheless, the apparent lineshape from these models continues to exceed the intrinsic quasiparticle scattering rate $\Sigma$ out to very large values of $\lambda$. This exemplifies the importance of the relative energy scales of $t_{\perp}$ and $\Sigma$, as a comparatively large $k_z$ dispersion is manifest as an artificially large quasiparticle scattering rate. This can become very important in connection to multi-orbital systems, where bands of different orbital character, and implicitly different $t_{\perp}$ may have the appearance of orbital-specific correlation strengths \cite{Day_2021}. We come back to this issue of relative energy scales below, when we consider extraction of band velocity and $k_F$ in greater detail.

The comparative success of the slab calculations in achieving high $k_z$ fidelity is encouraging, as it suggests the $I_{FT}$ description of $k_z$ integration underestimates our ability to achieve good $k_z$ precision experimentally. It remains unclear however, how this comes about, and what physical mechanism is missing from the bulk Fourier transform description of $k_z$ integration.

Before coming to a justification for this discord, we provide further evidence for the superiority of the slab methodology. To do so, we direct our attention towards the band centre, at $k_z = \pi/2c$ so as to avoid complications associated with asymmetries near the band edge. For our bulk Hamiltonian:
\begin{equation}
\epsilon(k_{||}=0,k_z) = -2t \mathrm{cos}(k_z),
\end{equation}
which can be inverted as
\begin{equation}\label{eq:mdc}
k_z = \mathrm{cos}^{-1}(-\epsilon/2t).
\end{equation}

\begin{figure}
\includegraphics[width=\columnwidth]{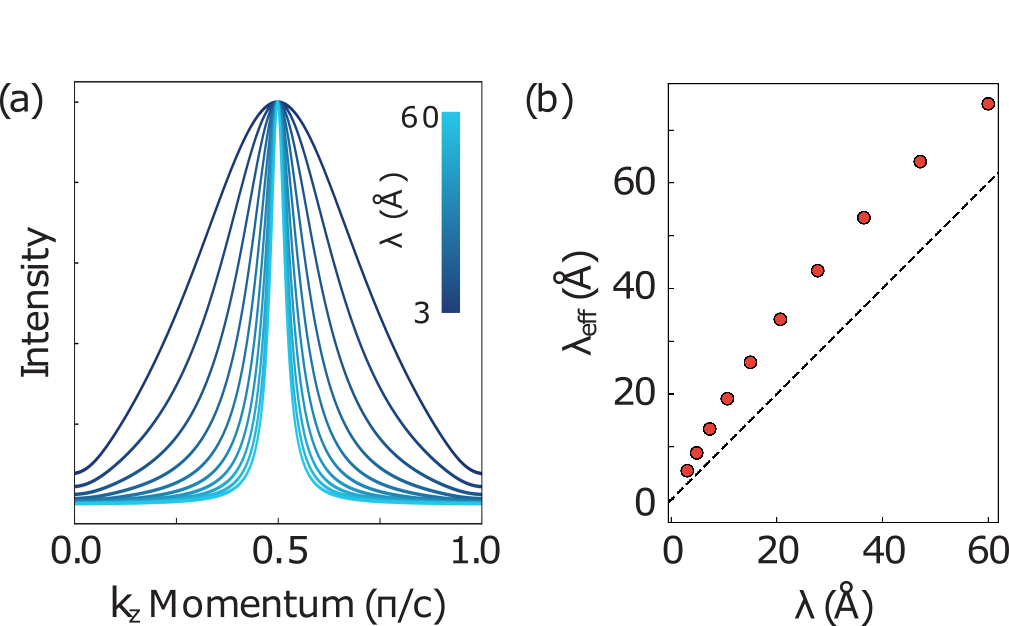}
\caption{ Spectral EDCs from Fig. \ref{fig:spectra}(d), converted to MDCs via inversion of bulk band dispersion from Eq. \ref{eq:mdc} in (a). The resulting MDCs follow a precise Lorentzian form, from which the width can be extracted, and converted to an effective $\lambda_{eff}$ utilizing Eq. \ref{eq:I_FT}. In (b), $\lambda_{eff}$ is plot against $\lambda$, the escape depth utilized in the simulation. The relation $\lambda_{eff}=\lambda$ is provided by the dashed black line.}
\label{fig:MDC}
\end{figure}

\begin{figure*}
\includegraphics[width=\textwidth]{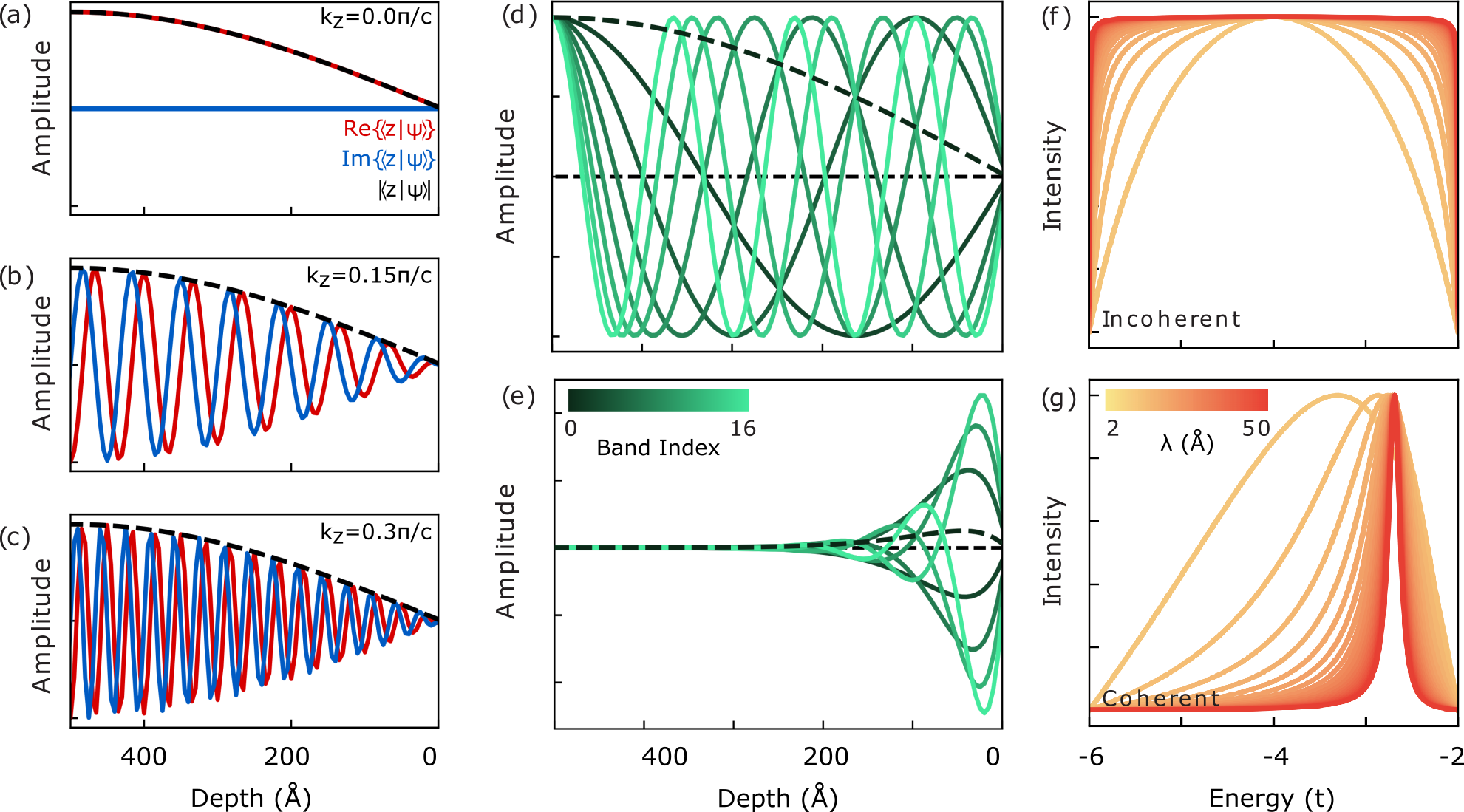}
\caption{Interlayer Interference: Wavefunctions for the state $|\psi_{\epsilon_{min}}\rangle$ at the band edge are plot for $k_z=0$ (a), $k_z=0.5\pi/c$ (b), and $k_z=\pi/c$(c), illustrating the rapid evolution of interlayer relative-phase with $k_z$. In (d), we plot the envelope $|\langle \phi_{\alpha}|\psi\rangle|$ for various states near $|\psi_{\epsilon_{min}}\rangle$. Colour increases from dark to light green with increasing band-index. The same envelopes, attenuated by an exponential with characteristic width $\lambda=100\ang$  are provided in (e). In (f,g) we compare the spectral feature at $k_z=0.8\pi/c$ with (f) and without (g) interlayer interference included. Escape depth $\lambda$ increases from yellow to red.}
\label{fig:envelope}
\end{figure*}
The simple form of this band dispersion enables an explicit mapping of energy distribution curves (EDCs) to momentum units (MDCs) in $k_z$. Doing so, the irregular lineshape of the EDCs is transformed to a Lorentzian function over $k_z$. In the limit of $\Sigma\rightarrow 0$, these MDCs allow a quantitative evaluation of the $k_z$ distribution of our slab model via the Lorentzian width. Moreover, in comparison with Eq. \ref{eq:I_FT}, this peak width can be inverted to extract an effective escape depth $\lambda_{eff}\equiv1/\Sigma$. The result, plot in Fig. \ref{fig:MDC} demonstrates that the slab-formalism is characterized by an effective increase in the photoelectron escape depth. Such an apparent increase of $\lambda$ coincides with reduced certainty in our position information, leading directly to an improvement in our $k_z$ fidelity as probed via ARPES. Not only do the slab calculations suggest better $k_z$ resolution, but we conclude that in general, surface sensitivity in ARPES cannot be reduced to a Lorentzian $k_z$ distribution. This is particularly apparent near the band edge, where such an interpretation fails to register the apparent band renormalization which follows from the reduced atomic coordination of the crystal's surface layers.

\section{Momentum Fidelity From Interference}

In the previous section, we observed that the slab construction supports results more consistent with experimental reality, where high $k_z$-fidelity is often achieved in materials with substantive $t_{\perp}$. The mechanism which underlies this discrepancy is related to photoelectron interference, similar to what is observed in graphene \cite{Chiang2011}, and other materials  \cite{WKu2012,Zhu2013}. We rewrite Eq. \ref{PES_spectral}:
\begin{equation}
I(k,\omega) = \sum_{i=1}^{mN} |M_{i,k}(\omega)|^{2}A_{i,f}(k,\omega),
\end{equation}
where the sum is over all eigenstates, and
\begin{equation}\label{eq:matel}
M_{i,k}(\omega) = \sum_{\alpha}c_{i,\alpha} \langle e^{ik\cdot r}|e^{-z_{\alpha}/2\lambda}\Delta|\phi_{\alpha}\rangle,
\end{equation}
is the photoemission matrix element. The sum over $\alpha$ extends over all basis states, with  $c_{i,\alpha} =\langle \phi_{\alpha} |\psi_i\rangle$. In our simple cubic tight-binding model, the $\alpha$ index correlates directly to the layers in the slab, such that we can use basis and layer indexing interchangeably. In more complex models like the one we use in Ref. \onlinecite{Day_2021}, each layer may include several basis states. As illustrated by the wavefunctions plot in Fig. \ref{fig:envelope}(a-c), for a given band $\psi_i$, the relative phase between projections onto different basis layers $\phi_{\alpha}$ varies rapidly with both depth and $k_z$. For the band edge state, the basis projections are in phase at $k_z=0$, as expected from the bulk model [$\epsilon_{bulk}(k_z=0)=\mathrm{min}\{\epsilon_{slab}\}$]. This leads to constructive interference, and finite signal from this band. This becomes destructive interference for $k_z\neq 0$, with total extinction of the signal at $k_z=\pi/c$. 

This is precisely the mechanism which enables the photoemission experiment to probe specific states within the slab's spectral continuum at a given photon energy: the oscillation in the wavefunction amplitude between neighbouring layers results in destructive interference for photoemission from states with $\langle P_+\rangle<1$. However, this is compromised upon introduction of surface sensitivity. The failure of perfect destructive interference is visualized in Fig. \ref{fig:envelope}(d,e) where we plot the wavefunction amplitude for several quantum-well states near the band-edge at $k_z=0$. When the photoelectrons associated with neighbouring layers are in phase, we have constructive interference and finite photoemission signal. Conversely, when these contributions are out of phase, the signal will vanish. When we account for surface attenuation, slowly oscillating amplitudes can produce finite signal, as the destructive interference becomes incomplete. This gives us a spectral broadening, or $k_z$ integration. Eventually, as we move away from the state with $\langle P_+\rangle=1$, the spatial frequencies increase to a point where destructive interference is largely recovered.

To unambiguously demonstrate the importance of interference here, we can reduce the sum over $\alpha$ in Eq. \ref{eq:matel} to an incoherent sum over all basis states. The results can be contrast in Fig. \ref{fig:envelope}(f,g). Without interference, the photoemission intensity becomes a flat distribution which falls off at the band edge with the inverse of the density of states. Restoring the coherent sum in Fig. \ref{fig:envelope}(f), we recover the familiar lineshape of the experimental photoemission spectra, peaked in energy at the bulk eigenvalue $\epsilon_{bulk}(k_z)$ in the large $\lambda$ limit. As a final note, the spatial distributions presented in Fig. \ref{fig:envelope}(e) provide some justification for the enhanced $\lambda_{eff}$ we saw previously. As the attenuated spatial distributions have widths $\Delta z>\lambda$, the effective position measurement is over a considerably wider depth than the exponential half-width.\textit{ We conclude from this that both $k_z$ selection, and integration are a direct consequence of photoelectron interference.} While the thin slice of material probed in photoemission is a radical departure from the infinite periodic lattice upon which the concepts of $k_z$-quantum numbers are constructed, the relative phases encoded in the wavefunctions provide the opportunity to resolve the $k_z$-dependent electronic structure with unprecedented success.

\section{Band Mapping at the Surface}
\begin{figure}
\includegraphics[width=\columnwidth]{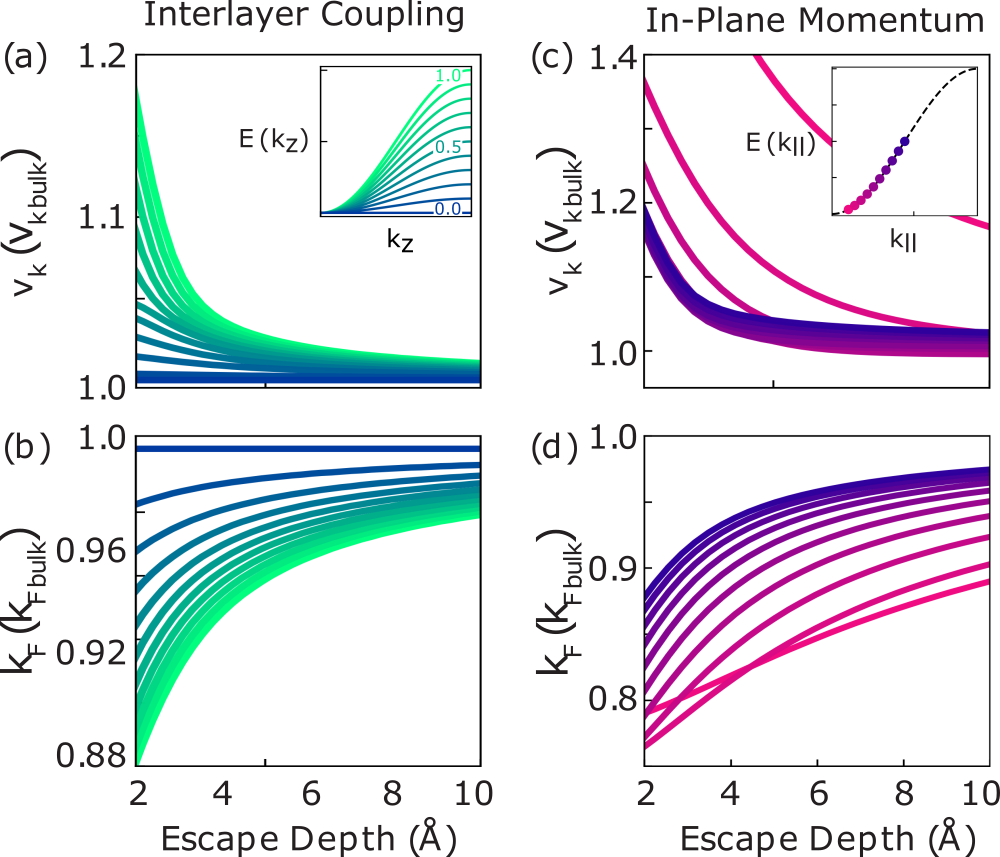}
\caption{ Band mapping at the surface. In panel (a), we plot the band velocity $v_k=\frac{\partial \epsilon}{\partial k}$ for various values of $t_{\perp}$ ($t_{||}=-1.0$) over a range of relevant $\lambda$ values. The inset indicates the $k_z$ band dispersion of the bulk model for each curve with different $t_{\perp}$ in the main figure. The corresponding plot for $k_F$ is provided in (b). The vertical axes are expressed in units of the bulk band velocity and bulk Fermi momentum, respectively. In (c,d), we repeat this analysis for a fixed value of $t_{\perp}=t_{||}$, at different values of in plane-momentum $k_{||}$. The in-plane momenta selected are indicated alongside the band dispersion of the bulk model in the inset of (c).}
\label{fig:tperp}
\end{figure}

Above, we demonstrated that the slab treatment utilized in the present work reflects experimental reality more reliably than the conventional heuristic. However, the fact remains that photoemission in the small $\lambda$ limit presents challenges for interpretation of the signal and its relationship with the bulk electronic structure. To provide a comprehensive perspective on the experimental consequences of this surface sensitivity, we study extraction of both the Fermi surface $k_F$ and band velocity $v_k=\frac{\partial E}{\partial k}$, as one would do for an experiment. In Fig. \ref{fig:tperp}(a), we plot the band velocity $\frac{\partial E}{\partial k}$ extracted from $E_F$ for the cubic model at half-filling. The spectral peaks fit well to a linear function, the slope of which is $v_k$. We repeat this process for several values of the out-of-plane hopping strength, $t_{\perp}$. In the small $\lambda$ limit, the band-velocity extracted can differ by anywhere from 10 $\%$ to 120 $\%$ of the bulk value, depending on the strength of interlayer hopping in the material. This converges to the bulk limit like $1/\lambda^2$. Estimation of the Fermi momentum is also influenced by this surface sensitivity. In Fig. \ref{fig:tperp}(b), we plot the analogous results for $k_F$ as a function of both $\lambda$ and $t_{\perp}$, illustrating the similar way in which band mapping is compromised in the presence of finite $k_z$ dispersion. While convergence is certainly slower for $k_F$, the absolute deviation is much less pronounced, with an absolute upper bound of approximately 10$\%$ error between the photoemission- and bulk- derived Fermi momentum. \textit{This more modest renormalization of the apparent $k_F$ suggests ARPES to be a more reliable measure of the bulk Fermi surface than is often assumed. }

This being said, while Fig. \ref{fig:tperp} may give the impression that surface sensitivity can be disregarded for $t_{\perp}<0.1t_{||}$, this is not a general rule, as the effects vary in severity across the bandwidth. This is made rigourous by repeating our analysis of $v_k$ and $k_F$ at different in-plane momentum $k_{||}$, again as a function of $\lambda$. Keeping $t_{\perp}$ fixed and equal to $t_{||}$, we observe that the extracted $v_k$ and $k_F$ become less reliable as one moves towards the band edge, as would occur when mapping the Fermi surface of a very shallow pocket. For large $t_{\perp}$, at the band edge one may very easily record a Fermi surface area significantly different from the bulk value. 

Nevertheless, we observe here that as a material becomes more two-dimensional, it becomes increasingly safe to disregard the capacity for $t_{\perp}$ and $\lambda$ to misdirect one's interpretation of band-mapping experiments.  Ultimately, as with many questions in physics, the importance of the $k_z$ uncertainty inherent to photoemission is a question of relative energy scales. The slab formalism we have detailed in this present work, made convenient via software tools like $chinook$, provides an excellent methodology by which such questions can be addressed \cite{Day2019}.

\section{Conclusion}
In this discussion, we have articulated the inadequacies of the canonical expressions by which $k_z$ integration and uncertainty are described in the context of angle-resolved photoemission spectroscopy. By extending consideration from the bulk electronic structure of an infinite lattice to a finite vacuum-terminated slab geometry, we have identified an interference mechanism encoded in the electronic eigenspectrum which results in improved $k_z$ fidelity, in line with experimental realities. Despite the acute surface sensitivity endemic to ultraviolet photoemission spectroscopy, this method provides a reliable measure of the three-dimensional electronic structure of solids well into the small escape depth limit. Nonetheless, the pernicious effects of finite $k_z$-uncertainty remain, and act to complicate estimation of both band dispersions and quasiparticle scattering rates. In regards to these experimental objectives, one must always account for the relative energy scales of both the $k_z$ band dispersions, and those one intends to resolve. In some cases, photon-energy dependent experiments over a broad energy domain can be utilized to establish convergence with the bulk limit. Ultimately however, one must really balance expectations of what information is sought, and what can realistically be accessed in the context of a given experimental configuration and material. 

\section{Appendix}


In the main text, we have assumed validity of periodic boundary conditions, effectively creating an infinite stack of floating crystalline slabs. One would expect that for a very thick vacuum layer, this construct should converge to the more physically correct case of true open boundary conditions. In our PBC model, we have taken a specific choice of gauge, wherein the $k$-dependent phase information is encoded in the initial state wavefunctions $\psi_i=\sum_{\alpha}c_{i,\alpha}\phi_{\alpha}$. This allows the simplification that the final state $e^{ik\cdot r}$ utilized in the evaluation of $\langle e^{ik\cdot r}|\Delta |\phi_{\alpha}\rangle$ can be performed over the local coordinates of the basis states $\phi_{\alpha}$. This phase information includes $k_z$-dependence, on account of the PBC utilized in the slab calculation. Going to OBC, the wavefunctions $\psi_i$ are no longer functions of $k_z$. However, the photoelectron final states are, and emanate from sites localized on different layers of the crystal. Consequently, a factor of $e^{ik_z z}$ still appears in the matrix element computed with open boundary conditions:
\begin{equation}
M_{i}^{OBC}(k) = \sum_{i}\sum_{\alpha} c_{i,\alpha}(k_{||})e^{ik_z z_{\alpha}}\langle e^{i k \cdot r}|\Delta|\phi_{\alpha}\rangle.
\end{equation}
By moving from periodic to open boundary conditions along the surface normal, we simply shift $k_z$-dependent phase information from our initial to final state functions in estimation of the photoemission matrix element, making the two constructions entirely compatible.
\section{Acknowledgements}
This research was undertaken thanks in part to funding from the Max Planck-UBC-UTokyo Centre for Quantum Materials and the Canada First Research Excellence Fund, Quantum Materials and Future Technologies Program. This project is also funded by the Natural Sciences and Engineering Research Council of Canada (NSERC); the Canada Research Chairs Program (A.D.); Canada Foundation for Innovation (CFI); British Columbia Knowledge Development Fund (BCKDF); and the CIFAR Quantum Materials Program.
\bibliography{surface}
\end{document}